\documentclass[twocolumn,showpacs,preprintnumbers,amsmath,amssymb]{revtex4}

\usepackage{graphicx}
\usepackage{dcolumn}
\usepackage{bm}

\begin{document}

\preprint{APS/123-QED}

\topmargin 0pt

\title{Local community extraction in directed networks} 

\author{Xuemei Ning$^1$}
\author{Zhaoqi Liu$^2$}%
\author{Shihua Zhang$^2$}
\email{zsh@amss.ac.cn}


\affiliation{%
$^1$College of Science, Beijing Forestry University, Beijing 100083, China
$^2$National Center for Mathematics and Interdisciplinary Sciences,
Academy of Mathematics and Systems Science, Chinese Academy of
Sciences, Beijing 100190, China
}

%
%

\date{\today}

\begin{abstract}
Network is a simple but powerful representation of real-world
complex systems. Network community analysis has become an invaluable
tool to explore and reveal the internal organization of nodes.
However, only a few methods were directly designed for community-detection in
directed networks. In this article, we introduce the
concept of local community structure in directed networks and
provide a generic criterion to describe a local community with two
properties. We further propose a stochastic optimization algorithm
to rapidly detect a local community, which allows for uncovering the
directional modular characteristics in directed networks. Numerical
results show that the proposed method can resolve detailed
local communities with directional information and provide
more structural characteristics of directed networks than previous
methods.

\end{abstract}

\pacs{Valid PACS appear here}


\maketitle

\section{Introduction}
Networks consisting of nodes connected in pair by edges reveal
essential features of the structure, function and dynamics of many
complex systems. Thus, complex networks have become invaluable tools
in various fields including sociology, biology and physics
\cite{Newman2003,Zhang2007}. The characteristic of community structure in networks
can aid in exploring the structure and organization of networks.
In the past decade, it has attracted huge attentions. Many methods
for resolving community structure in undirected networks have been
developed (see Ref. \cite{Fortunato2010} for a recent comprehensive
review). However, only a limited number of methods were designed for
detecting community structure in directed networks and the direction
of links leads to new challenges in defining community structure of
directed networks \cite{Guimera2007,Leicht08,Kim2010}.

Directed networks show fundamentally different features
when the direction of their links are ignored. The link direction
characterizing important topological information is essential
to describe the structure of many complex systems. The effects of
link directions to the organization and dynamics of complex networks
have attracted great interests recently. For instances, link
direction has been proven to play profound effects on link tendency
between nodes \cite{Foster2010}. The studies on community detection
in directed networks have shown that considering link direction can
shed light on key structural features of community structure in
directed networks \cite{Guimera2007,Leicht08,Kim2010}.

How to describe the community structure in a directed network is an
open issue. Newman and Leicht \cite{Newman2007} and Guimera \emph{et
al.} \cite{Guimer2005} have defined a community that nodes are
assigned to it when they are linked to similar neighbors. This
definition of a community is fundamentally different from the
general one used for undirected networks
\cite{Fortunato2010}. Moreover, Rosvall and Bergstrom
\cite{Rosvall2007} have adopted an information theory-based method
which shows distinct characteristics with adapted modularity
maximization method. Leicht and Newman \cite{Leicht08} and Kim
\emph{et al.} \cite{Kim2010} have attempted to employ the
generalized form of modularity to identify the community structure
in directed networks, respectively. However, similar to modularity
partition methods for undirected networks, such type of methods
which force every node into a community can distort the real
structure of the networks, in which, some nodes may only loosely
connected to any community. Moreover, the modularity index
\cite{Newman2004,Newman2006} has been shown to fail to find the most
natural community structure in undirected networks due to the
resolution limit issue \cite{Fortunato2007,Li2008}, which would be
shared with the adapted modularity for directed networks.

More recently, the concept of local community was proposed for
undirected networks \cite{Zhao2011,Zhang2011}. The key idea is that,
in a large network, a community, focusing on the ``local'' links
within and connecting to it, refers to a limited number of nodes in
the whole network. The principle of determining such a local
community at a time is different from the partitioning methods,
which consider the whole connections of a network. There has been no
much work in the literature focusing on the local community
detection even for undirected networks. Researchers have explored a
community around a given node which relies on the predefined
knowledge \cite{Flake2002,Clauset2005}. Zhao \emph{et al.}
\cite{Zhao2011} proposed a community extraction framework
considering only one community at a time by maximizing an extraction
criterion via tabu search technique. The promising idea and the
issue of resolution limit of the proposed criterion have inspired a
neurodynamic framework with a generic criterion to resolve local
communities in undirected networks, recently \cite{Zhang2011}. Taking
into account the complexity of directionality and intricate
connections between nodes, we adopt the ``local" strategy to
disassemble and study the directed networks here.

\begin{figure*}[t]
\begin{center}
\includegraphics[width=1.00\textwidth]{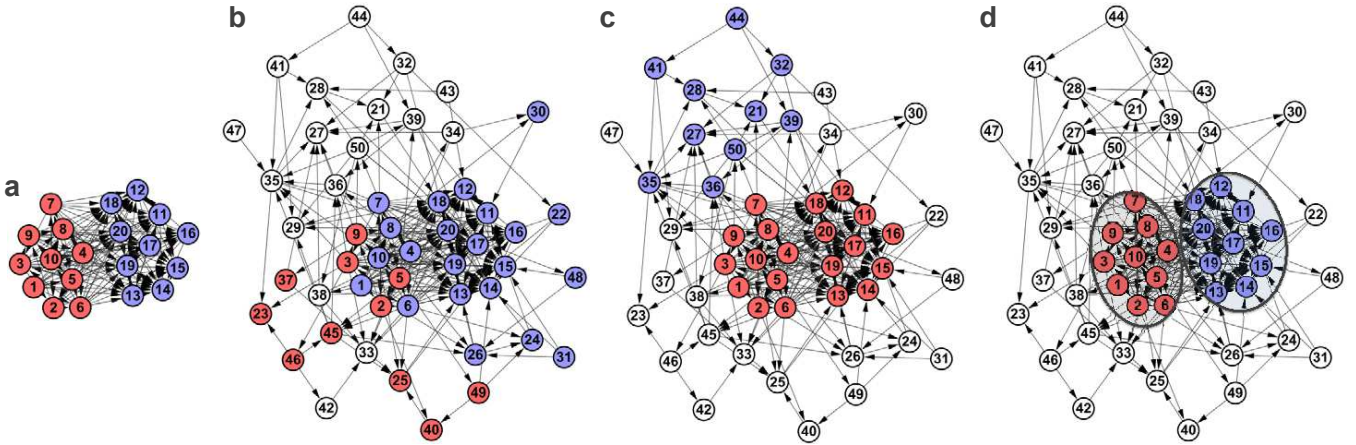}
\caption{Illustration of three methods for discovering community
structure in a directed network. The network consists of $50$
nodes, and the first 20 nodes belong to a dense subnet where links
between members form independently with probability 0.7. The links
between members and the other 30 nodes and links between the other
30 nodes all form independently with probability 0.1. We assign
directions to the links within the first 10 nodes, the second 10
nodes, and other 30 nodes randomly. While for links between the
first 10 nodes and other 40 nodes, all are assigned directions from
the first 10 pointing to other 40. As to the second 10 nodes, all
the links are assigned directions from other 40 pointing to them. A
partition into three communities using the directed modularity
maximization (DMM) by Leicht and Newman, the undirected community
extraction (UCE) without considering the directionality of the
network and our directed community extraction (DCE) method are shown
in (b), (c) and (d), respectively. Different colors represent the communities
detected by each method. The two circled regions in (d) represent the
two true communities respectively. If we only consider the 20 nodes
of the network by removing the 30 backgroup nodes, as stated by
Leicht and Newman, DMM can identify the two communities by a
partition with two communities (a). However, in the current network, DMM has to balance
tightness of the three communities, and as a result distort the
community structure (b). UCE can well extract the 20 nodes as a
dense community, but fail to detect the directed communities (c).
Our DCE method, on the other hand, separates out the true community
perfectly (d). } \label{figurecurve}
\end{center}
\end{figure*}
In this article, we introduce a generic quantitative criterion to
describe a ``local" community in directed networks (Figure 1). The
generic criterion considers two properties: (1) high density--the
sets of nodes in a community are densely connected; (2) consistent
directionality--the direction of links between a community and the
rest of networks should be as consistent as possible. We can see that finding sets of
nodes that optimize this measure is in general a computationally
challenging problem. We adopt a Markov chain Monte Carlo (MCMC)
approach to sample from sets of nodes according to a distribution.
This distribution gives significantly higher probability to sets of nodes with
high density and consistent directionality. MCMC is a
well-established technique to sample from combinatorial spaces with
applications in various fields \cite{Gilks1998,Randall2006}
including bioinformatics \cite{Bansal2008,Vandin2011}. In general,
the computation time (e.g., number of iterations) required for an MCMC
approach is unknown. In our case, we empirically show that our
MCMC-type algorithm converges rapidly to the stationary
distribution and it can scale well with respect to networks with
10000 nodes. Numerical results show that our local community
extraction method can resolve local communities with
directional information and provide more structural
characteristics of directed networks than previous methods.

\section{Methods}
\textbf{Local community extraction problem in directed networks} We
first introduce the local community extraction problem in undirected
networks. Let $G(V,E)$ denote an undirected network of $N$ nodes. The network
is denoted by a symmetric adjacency matrix $A=[A_{ij}]$ of
size $N\times N$, where $A_{ij}>0$ if there is an edge between nodes
$i$ and $j$ and $A_{ij} =0$ otherwise. The positive $A_{ij}$'s are the
weights for weighted networks; or they are set to 1 for unweighted networks.
The kernel idea of local community extraction problem is to look for
a set $S$ of nodes with a large number of links within itself and a
small number of links to the rest of the network. This problem can
be described to optimize a quantitative function. Note that the
links within the complement $S^c$ of this set do not affect the
value of this function. Recently, we have introduced a generic
quantitative criterion $W_S$ to describe local communities in
undirected network \cite{Zhang2011} which adapts the one proposed in
\cite{Zhao2011} with a parameter $\rho$. We note that the
generalized criterion can reveal multi-resolution community
structure and conquer the resolution limit issue of the previous
one. Specifically, it can be defined as follows,
\begin{equation}
W_S=|S||S^\rho|
\left[\frac{O_S}{|S|^2}-\frac{B_S}{|S||S^\rho|}\right], \label{qg2}
\end{equation}
where $|S^\rho|=\rho N-|S|$, $2\frac{|S|}{N}<\rho\leq 1$, and $O_S=
\sum\limits_{i,j\in S} A_{ij}$, $B_S=\sum\limits_{i\in S,j\in S^c}
A_{ij}$. The $|S^\rho|$ can be considered as the estimation of the
number of nodes connecting to the community $S$ in the rest of the
network. When $\rho=1$, it is the one proposed in \cite{Zhao2011}.
The term $O_S$ is twice the weight of the edges within $S$, and
$B_S$ denotes connections between $S$ and the rest of the network.
The maximization of $W_S$ can be solved efficiently by a powerful
neurodynamic framework.

Now we consider a ``community'' in a directed network
$\overline{G}(V,E)$. The network can be represented by an asymmetric
adjacency matrix $A=[A_{ij}]$, where $A_{ij}>0$ if there is an edge
directed from node $i$ to node $j$, $A_{ij}=0$
otherwise. The key point is that the community structure should
reflect the ``directionality'' in the directed network.
The above criterion have well considered the density of a community
and sparse connections to the rest of the network. Here we
incorporate a parametric coefficient to capture the potential effect
of directions,
\begin{equation}
W_S^d=|S||S^\rho|
\left[\frac{O_S}{|S|^2}-q_d^n\frac{B_S}{|S||S^\rho|}\right],
\label{qg2}
\end{equation}
where $q_d=\frac{B_S+1}{|B_S^{in}-B_S^{out}|+1}$ and $O_S=
\sum\limits_{i,j\in S} A_{ij}$, $B_S=B_S^{in}+B_S^{out}$,
$B_S^{out}=\sum\limits_{i\in S,j\in S^c} A_{ij}$,
$B_S^{in}=\sum\limits_{i\in S^c,j\in S} A_{ij}$ and $1\leq q_d\leq
B_S+1$. If the directions of all links between $S$ and $S^c$ are
consistent, i.e., $B_S^{in}=0$ or $B_S^{out}=0$, then $q_d$ is equal to
1 which has no effect to the criterion $W_S^d$. While if the
directions are inconsistent,  the $q_d$ gets larger than 1 that
penalizes the second term. Note that $n$ is a parameter to control
the degree of penalty.

We note that the problem to maximize the generic criterion $W_S^d$
is computationally difficult, and it is likely that there is no
efficient algorithm to solve this problem exactly. Although the
neurodynamic framework has shown well performance, it is not
directly applicable to the current problem due to the effect of the
multiplier $q_d^n$. We consider to develop a stochastic search
procedure to solve this problem.

\textbf{A MCMC approach} We introduce a MCMC approach to solve the
problem described above. The MCMC approach samples sets of nodes, with the
probability of sampling a node set $S$ proportional to the objective
weight $W_S^d$ of the set. Thus, the frequencies that node sets are
sampled in the MCMC method provides a ranking of node sets, in which
the sets are ordered by decreasing sampling frequency. Thus, in
addition to the highest objective weight set, one may also examine
other sets with high objective value (``suboptimal'' sets) that are
nevertheless significant.
\begin{figure*}[t]
\begin{center}
\includegraphics[width=0.95\textwidth]{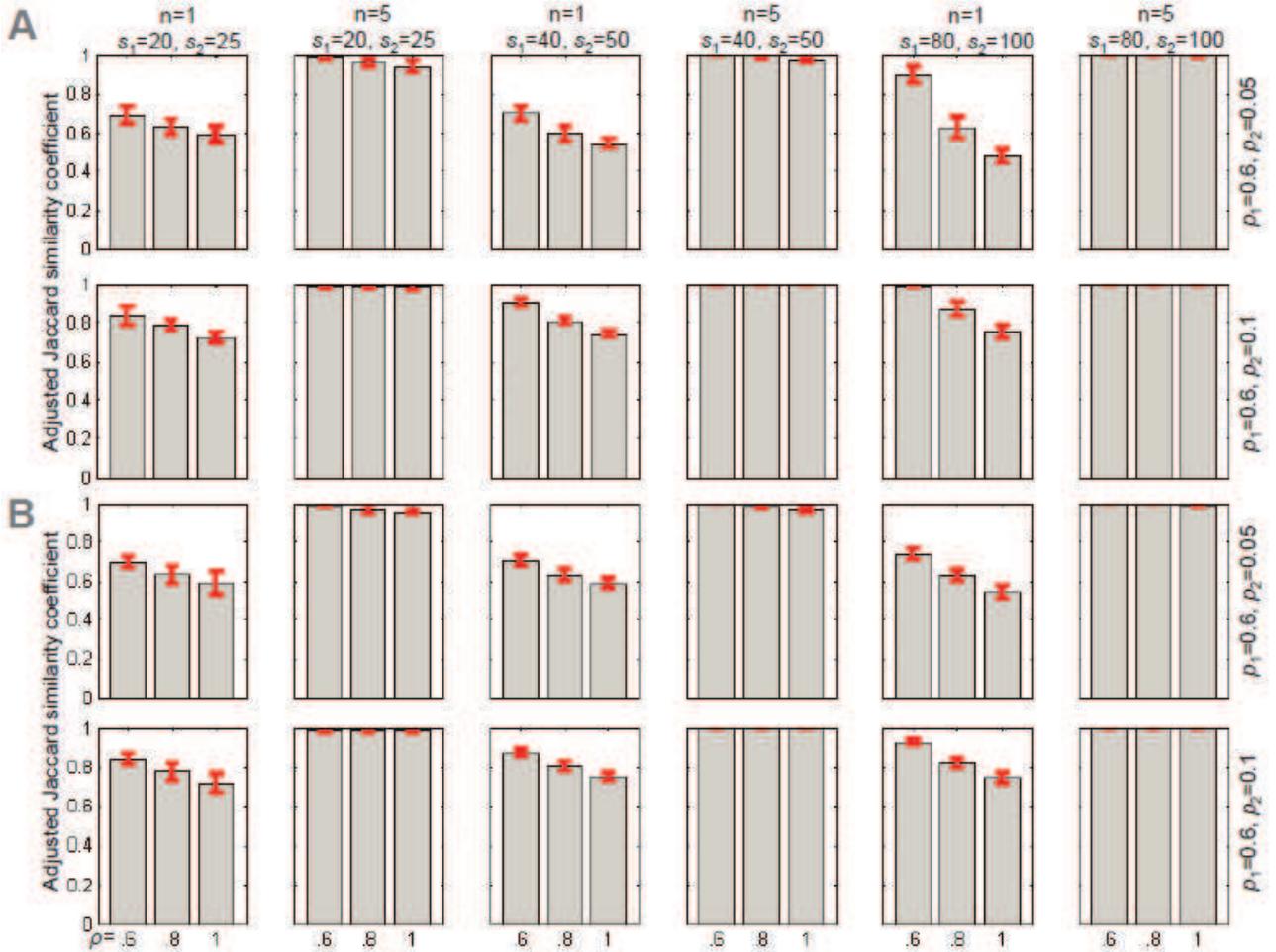}
\caption{Tests of local directed modularity optimization on the
benchmark via bar plots of adjusted Jaccard similarity coefficient measure.
The number of nodes $N=500$ for (A) and $N=1000$ for (B) respectively. Two
communities of three different sizes were embedded with two different
parameter settings $p_1$ and $p_2$. Different parameters $\rho$ and $n$ were
tested and shown. Each bar plot corresponds to an average over 20 network realizations.}
\label{figure.toyexample}
\end{center}
\end{figure*}

The basic idea of the MCMC is to build a Markov chain whose states
are the subsets of nodes of the network and to define transitions
between the states that differ by one node. The Metropolis-Hastings
algorithm provides a general method for designing transition probabilities
that gives a desired stationary distribution on the state space.
However, the Metropolis-Hastings method does not guarantee fast
convergence of the chain, which is a necessary condition for
practical use of this method. In fact, if the chain converges
slowly, it may take an impractically long time before the chain
samples from the desired distribution. Defining transition
probabilities so that the chain converges rapidly to the stationary
distribution remains a challenging and important task in real
applications. Despite significant progress in recent years in
developing mathematical tools for analyzing the convergence time
\cite{Randall2006}, our ability to analyze useful chains is still
limited, and in practice, most MCMC algorithms rely on simulations
to provide evidence of convergence to stationarity \cite{Gilks1998}.

We devise a Metropolis-Hastings algorithm to sample sets $S_G$ of
nodes with a stationary distribution that is proportional to
$e^{cW(S)}$ for some $c > 0$. At time $t$, the Markov
chain in state $S_t$ chooses a node $u$ in the neighborhood of
$S_t$, and moves to the new state $S_{t+1} = S_t\setminus\{u\}$ or
$S_{t+1} = S_t\bigcup\{u\}$ with a certain probability. In general,
there are no guarantees on the rate of convergence of the
Metropolis-Hasting algorithm to the stationary distribution.
However, we empirically demonstrate that in our case the MCMC
rapidly converges, and thus the stationary
distribution of a ``local" subnet is reached in a practical number
of steps by our method.
\\ \\
\textbf{Algorithmic procedure} \\
\textbf{Initialization:} Choose an arbitrary small
subset $S_0$ of nodes in $G$ (the set of all nodes). \\
\textbf{Iteration:} For $t = 1, 2,...,$ obtain $S_{t+1}$ from $S_t$
as follows:                                    \\
\textbf{1} Choose a node $u$ uniformly at random from $S_t^c$ (it is the
closure of $S_t$, i.e., $S_t\bigcup\{\mbox{all neighbor nodes of } S_t\}$).                             \\
\textbf{2} If $u\in S_t$, let
$P(S_t,u)=$min$[1,e^{cW(S_t\backslash{u})-cW(S_t)}]$; With
probability $P(S_t,u)$ set $S_{t+1}=S_t\backslash{u}$, else
$S_{t+1} = S_t$.                               \\
\textbf{3} If $u\in S_t^c\backslash S_t$, let $P(S_t,u)=$min$[1,
e^{cW(S_t\bigcup\{u\})-cW(S_t)}]$; With probability $P(S_t,u)$ set
$S_{t+1} =S_t\bigcup\{u\}$, else $S_{t+1} = S_t$.

The MCMC method is very promising and efficient due to
the speed of convergence of the Markov chain to its ``local'' stationary
distributions. We have shown that our method can scale well with large-scale
network of 10000 nodes.

\textbf{Stop criterion} After determining a local community, our
method can be further applied to its complement in the network to
extract the next community. How to determine the number of local
communities in a network is a hard, but practically important problem. In real
applications, we would suggest to evaluate the statistical
significance of a community by comparing its objective value with
that of 100 random directed networks generated by reserving the same
set of nodes and the same number of edges \cite{Maslov2002}.

\section{Results}
\begin{figure}[t]
\begin{center}
\includegraphics[width=0.48\textwidth]{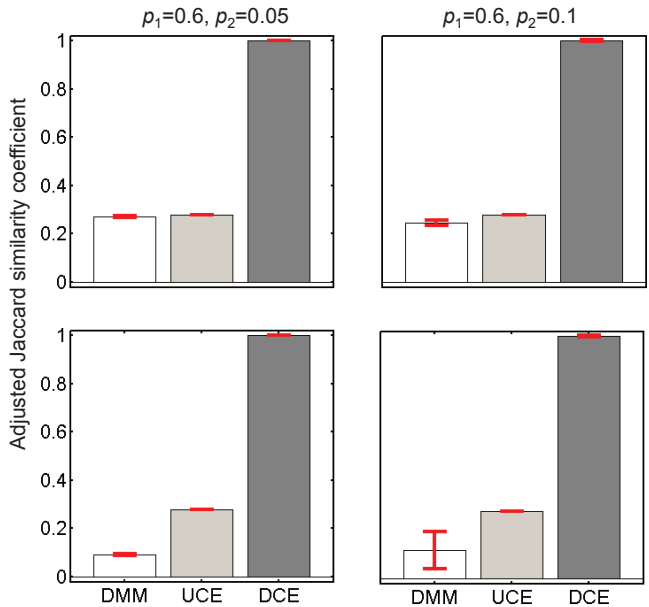}
\caption{Comparison of three methods via bar plots of adjusted Jaccard
similarity coefficient. The three methods are the directed modularity
maximization (DMM), the local undirected community extraction (UCE) method
and our method (DCE) in this paper respectively. The number of nodes $N=500$
for top two subfigures and $N=1000$ for bottom subfigures. Two communities
with sizes of $S_1=40$ and $S_2=50$ are embedded in the simulated networks.
Two types of connections are tested based on $p_1$ and $p_2$. Each bar plot
corresponds to 20 network realizations.} \label{figure.toyexample}
\end{center}
\end{figure}
\textbf{Numerical tests} We first test the directed community
extraction (DCE) criterion $W_S^d$ maximized by the MCMC algorithm and
further compare it to the undirected community extraction (UCE) criterion
$W_S$ ignoring the link direction \cite{Zhao2011}, and the
generalized directed modularity maximization (DMM) method proposed by
Leicht and Newman \cite{Leicht08} on simulated directed networks. To
compare grouping results against the independent partitions defined
by the embedding communities, the adjusted Jaccard similarity coefficient
as a measure of agreement is used for assessments. The Jaccard similarity
coefficient is defined as the size of the intersection divided by the
size of the union of the two sets:
$$J(A,B)=\frac{|A\bigcap B|}{|A\bigcup B|}.$$

We simulate a directed network of $n_{12}+n_0$ nodes starting with a
set $S_{12}$ of $n_{12}$ densely connected nodes and weakly connected
background $S_0$ of $n_0$ nodes. Each pair of nodes in $S_{12}$ and
$S_0$ are connected by links independently and uniformly at random
with probability $p_1$ and $p_2$. The direction of links within $S_{12}$ and $S_0$
are assigned at random but for links that fall between a subset
$S_2$ of $n_2$ nodes in $S_{12}$ and others are assigned directions
from $S_2$ to others. While for links between $S_1$($S_{12}\backslash S_2$) and others are
assigned direction at random from others to $S_1$
($S_{12}\backslash S_2$) (see Figure 1 for an example).

Given a result of two communities $C_i$ ($i=1,2$) generated by a method, we adopted the
following definition of adjusted Jaccard similarity coefficient to measure
the accuracy in our simulation study:
$$J(S,C)=\mbox{max}_{i,j\in \{1,2\}, i\neq j} \frac{1}{2}\left(\frac{S_1\bigcap C_i}{S_1\bigcup C_i}+\frac{S_2\bigcap C_j}{S_2\bigcup C_j}\right).$$
It is the degree-normalized maximum of all the possible sums of Jaccard
similarity coefficient of two groups of local communities. When the measure
is equal to 1, it implies that the two true groupings are perfectly
identified by the tested method.

We first apply our method onto various networks with different
connection characteristics (determined by parameters $p_1$ and
$p_2$) and test the effect of different criterion parameters (i.e.,
$\rho$ and $n$). We extract the first two communities by our method
for the calculation of the adjusted Jaccard similarity coefficient.
The results clearly depend on parameters $p_1$ and
$p_2$ of the benchmark, and parameters $\rho$ and $n$ of the
proposed directed local modularity criterion $W_S^d$ (Figure 2).

We can see that the results are more accurate with $n=5$ than those with $n=1$,
indicating that the parameter to control the degree of penalty is helpful.
We can also see that the results with $\rho=0.6$ are better than those with $\rho=1$,
suggesting that the original quantitative function use the number of all
complementary nodes of a community is problematic in some cases. In the following, we will
choose $\rho=0.8$ and $n=5$ for further comparative analysis.

\begin{figure}[t]
\begin{center}
\includegraphics[width=0.47\textwidth]{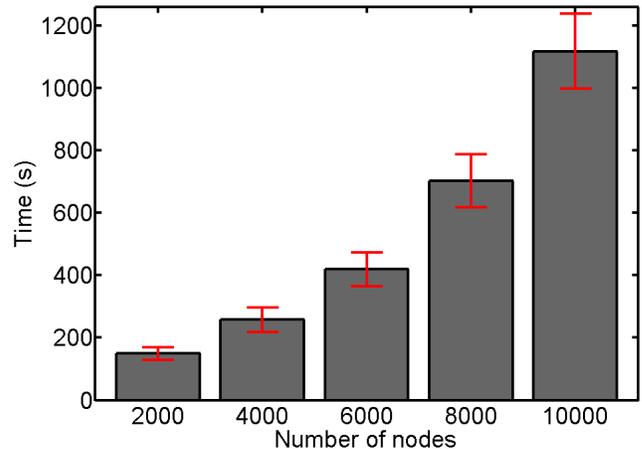}
\caption{Testing the running time of our method for one community with respect to
the network size from $N=2000$ to 10000. Each bar corresponds to an average
over 20 network realizations.} \label{figure.toyexample}
\end{center}
\end{figure}

We further compare our method with the other two methods. For a fair
comparison, we extract two communities by our method and the
undirected local community extraction method respectively, and we partition the network into
three parts by the directed modularity maximization method to allow one for
background nodes (Figure 3). We can clearly see that our method performs
the best for all four settings. While undirected community extraction
usually merge the two directed communities as one community and
extract another ``dense'' subset as its second community. Directed
modularity maximization improves slightly for denser communities,
but it tends to add the background nodes to a community, resulting
in poor overall adjusted Jaccard similarity coefficient. For
large-scale networks, this situation even
gets worse due to the resolution limit of modularity-type of
methods. Actually, even for small-scale networks with only 50 nodes,
we can see that the directed modularity maximization can not
identify the embedded communities well (Figure 1b). This is
partially because the connectivity within the background, and
between it and real communities can affect the (directed)
modularity. If we remove the background nodes and links, the
directed modularity method can identify the two communities (Figure
1a). All these results show that ``local'' extraction strategy reduces
the effect of background nodes, and improves the performance of
``partition'' type of community detection methods.

The computational efficiency of the proposed method can also be
seen in the simulation study, where we have applied our method
onto networks with 10000 nodes. The experimental analysis
have shown that our method can scale well (Figure 4).

\begin{figure*}[t]
\begin{center}
\includegraphics[width=0.8\textwidth]{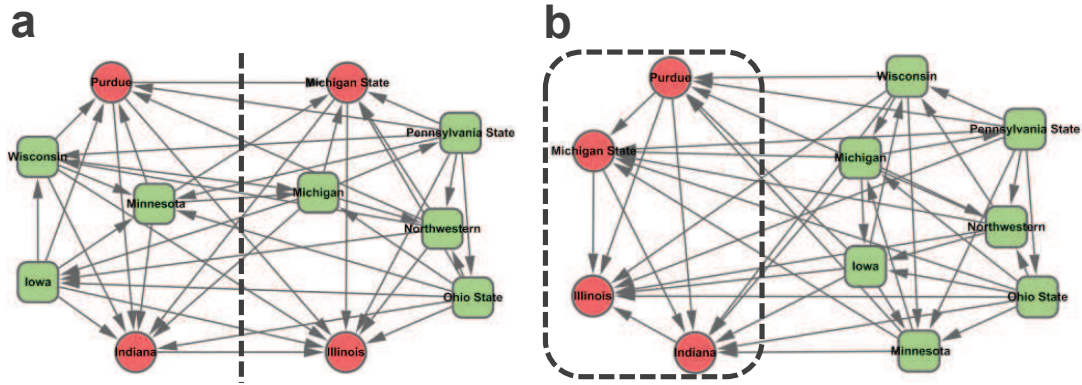}
\caption{Community detection in the small football network.
Two communities separated by the vertical dashed line are identified
by the undirected modularity maximization (a). And the box covered
nodes represented the first community extracted by the local directed extraction
method (b). The box covered region in (b)
represent the community, in which all teams lost a majority of their
games.} \label{figure.toyexample}
\end{center}
\end{figure*}

\textbf{Real applications} We further apply our method onto a
directed sporting competition network of US universities in the
American football game during the 2005 season which was firstly used
by Leicht and Newman recently (Figure 5) \cite{Leicht08}. The nodes
represent the teams in the `Big Ten' regional competitions or
`conference', and the edges link pairs of teams that played one another.
The direction of each edge reflects the win or lose relationship
between the two competing teams, i.e., the edges pointing from the winner
to the loser of each game. The traditional representation is
undirected which may miss important information. Our method and
directed modularity maximization method can precisely extract a
community including four teams, in which all of them lost a
majority of their games. While the undirected community extraction
method (UCE) and undirected modularity maximization fail to
identify it. They only extract a community with five teams randomly
due to the symmetric connectivity property of all nodes. This small network
clearly demonstrates that the edge directions play vital roles in
forming the community structure of a network.

\begin{figure*}[t]
\begin{center}
\includegraphics[width=0.8\textwidth]{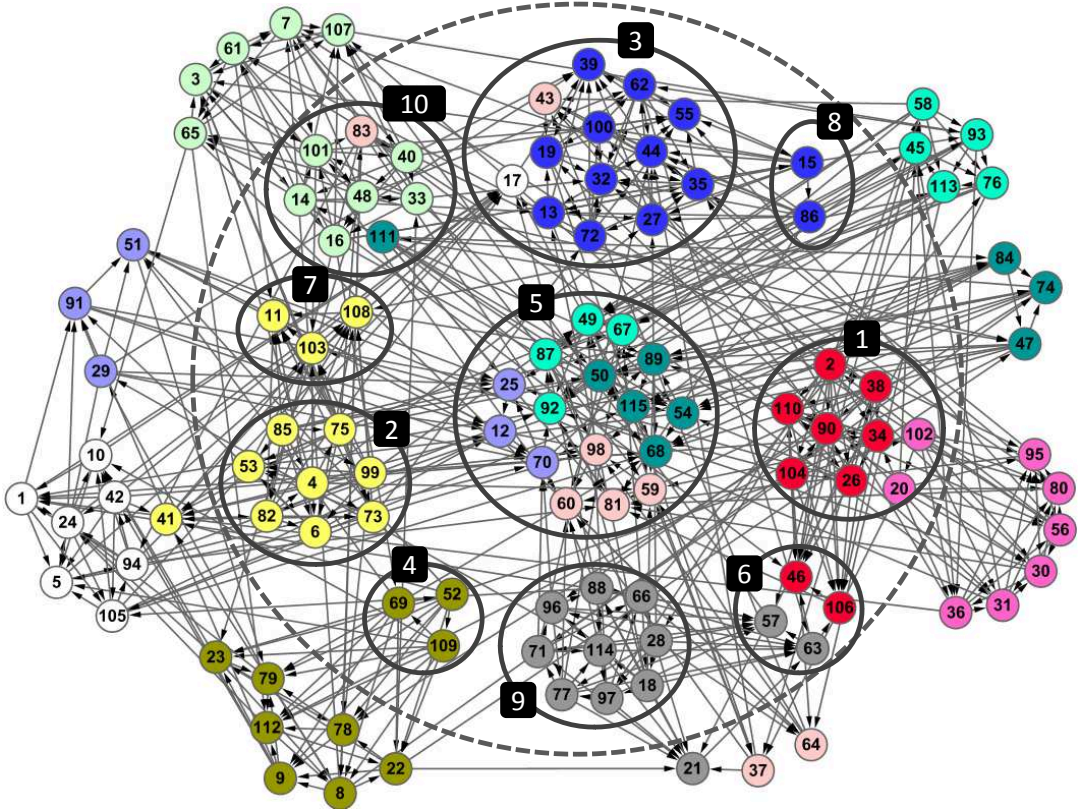}
\caption{Community identification in the directed football network by our method.
Colors represent the original 11 conferences and 8 independence
teams (soft red). The identified local communities were grouped in
circles and the corresponding number in the shaded box represent the
rank of their scores. We can see that the extracted region represent
the community, in which all teams lost or won a majority of their
games against all others.} \label{figure_football}
\end{center}
\end{figure*}

The small football network represents a regional conference (`Big
ten' conference) which likely corresponds to a community in the
undirected network of the whole country. We next apply our method
onto another directed football network of the whole country to show its
advantages with $\rho=1$ and $n=8$ (Figure 6). We should note that its
corresponding undirected version has been comprehensively used as a
gold testing system for evaluating the community-detection methods
in undirected networks. The football network originally compiled by
Girvan and Newman \cite{Newman2002} contains the competition
relationships of American football games between Division IA
colleges during regular season Fall 2000. The node and edge of this
network represent every team and every game played between two
teams respectively. Meanwhile, the nodes were marked with colors indicating the
conferences to which they belong. Note that the assignments to
conferences, the node colors, were corrected recently
\cite{Evans2010}. Here, we label the win and loss relationship
between two competing teams in this football network and construct a
directed football network to test our method.

Our method has shown very different community structure
with the original conferences (or computationally
community-detection in its undirected version).
We also have applied DMM to this network which has identified the similar
community structure with the DMM on
undirected version as previous tested. The DMM fails to capture the directional
information. While the DCE method discovers distinct community characteristics (Figure 6).
For example, the community 1 consisting of 8 teams, each of which won most of their
games with respect to all other teams. We may consider it be a strong group. While
community 3 failed most of their games, we may see it as a weak group. This community
structural organization format has revealed different properties compared to the
original conference organization. This exploration provide more insights into the
topological organization and enhance our understanding to the underlying principle of
this network.

\section{Conclusion}
How to describe community structure of directed networks is an open
issue in network science. It has attracted many people with broad
range of interests of diverse fields including physics, sociology,
biology and so on. In this article, we investigate the community structure problem in
directed networks from a ``local" view. We propose a new framework
for recovering the local community structure in directed networks by
optimizing a generic criterion via MCMC stochastic search
techniques. We further apply it to both simulated and real
networks to demonstrate that it is able to recover known local
community structure and reveal unexpected local patterns which can
not be recovered when ignoring the direction information.
The main purpose of this article is to propose the new
concept and theoretical framework to analyze the community structure
of directed networks which shed lights on the network's organization
and dynamics.


\begin{acknowledgments}
This work was supported by the National Natural Science Foundation of China, No. 61379092, 61422309 and 11131009, the Outstanding Young Scientist Program of CAS, the Scientific Research Foundation for ROCS, SEM, and the Key Laboratory of Random Complex Structures and Data Science at CAS. The authors thank Professor Mark Newman for providing the source code of directed modularity maximization method.
\end{acknowledgments}


\end{document}